\documentclass{article}
\usepackage[left=3cm,right=3cm,top=1.5cm,bottom=2cm]{geometry} 
\usepackage{amsmath} 
\usepackage{graphicx}
\usepackage{subcaption}
\usepackage{stackengine}
\usepackage{float}
\usepackage{amsthm}

\begin{document}

\newtheorem{defi}{Definition}
\newtheorem{theo}{Theorem}
\newtheorem{prop}{Proposition}
\newtheorem{lem}{Lemma}
\newtheorem{cor}{Corollary}
\newtheorem{refin}{Refinement}

\title{The bounded 15-vertex model}
\author{Kari Eloranta\\
  University of Helsinki\\
{\tt kari.v.eloranta@gmail.com}}

\date{\today}
\maketitle

\begin{abstract}
\noindent The 15-vertex model of Statistical Mechanics is studied on a square domain with partially oriented boundary. With DWBC the model would reduce to the six-vertex model, but more general boundary configurations are available. After establishing the dynamic version of the model we simulate with it to find the typical equilibrium states for a set of increasingly complex boundaries. Among others they yield almost isotropic non-trivial limit shapes even though the microscopic model is highly asymmetric.

\end{abstract}

\section{Introduction}

In the vertex models of Statistical Mechanics the spin variables at vertices are replaced by arrow orientations on the edges of a given lattice. A {\bf vertex rule} determines the allowed local vertex configurations and once these are obeyed everywhere globally we have a legal vertex configuration.

The best know such model is the six-vertex/Ice model defined on the square lattice (\cite{B}). Its vertex rule requires exactly two incoming and two outgoing arrows. This can of course be relaxed in various ways, some of them even physically meaningful. Perhaps the two best know in physics literature lead to 15-vertex and 19-vertex models (\cite{WM}). The latter posits as its vertex rule \lq\lq an equal number of incoming and outgoing arrows\ \rq\rq (unoriented edges are allowed). With direct enumeration one gets 19 allowed vertex configurations.

The 15-vertex rule is a simplification of the 19-vertex rule. Its allowed vertex configurations are illustrated in Figure 1 (unoriented edges dotted).

\setcounter{figure}{0}
\begin{figure}[H]
\centerline{\includegraphics[height=1.6cm]{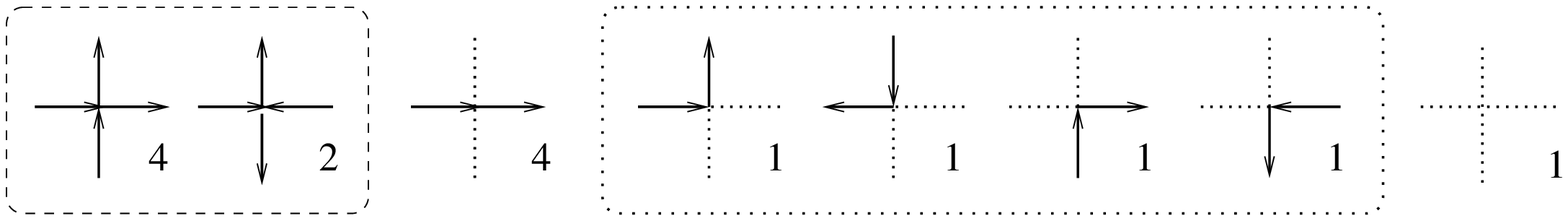} \hskip 1cm \includegraphics[height=1.6cm]{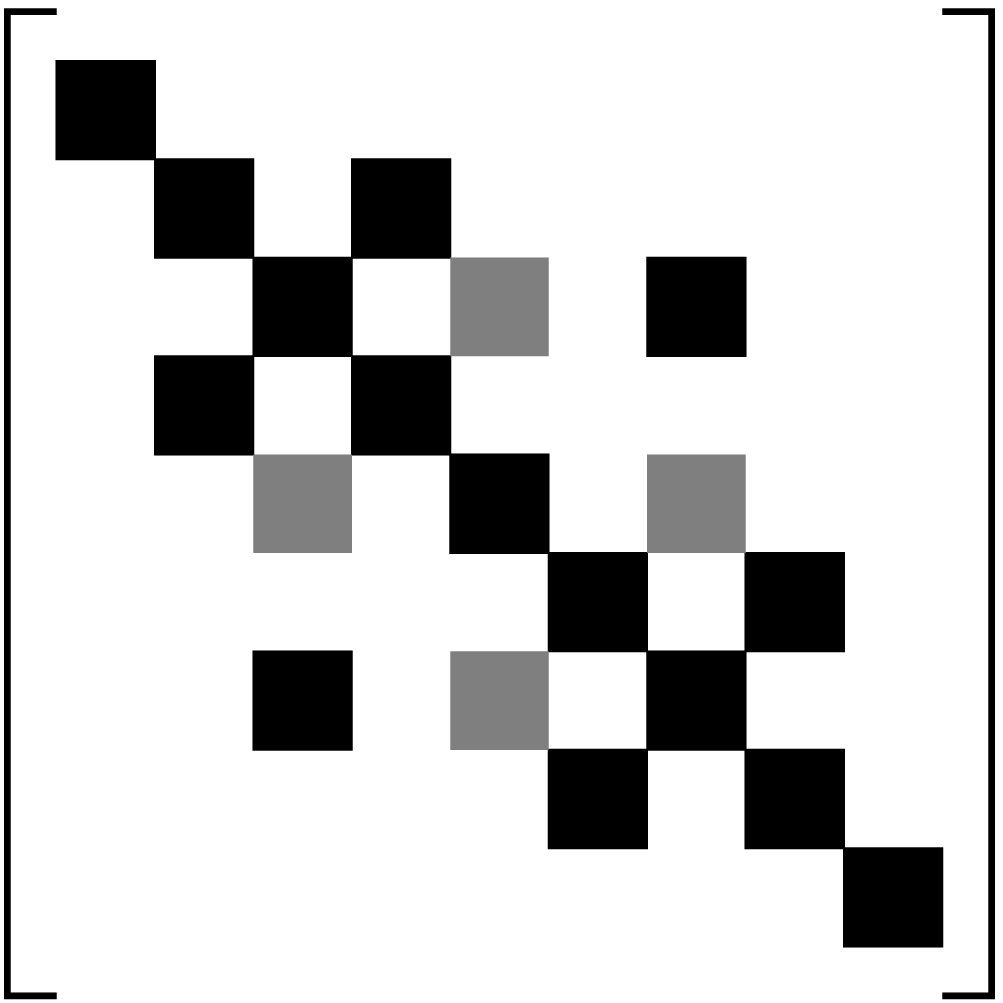}}
\caption{15-vertex rule: allowed vertex configurations with multiplicities. R-matrices.}
\end{figure}

\noindent Inside the broken line on the left is the doublet of allowed Ice vertices. The quadruplet inside the dotted frame in the center is the one distinguishing the 15-vertex rule. If the missing rotations of these \lq\lq L-type\rq\rq\ vertex configurations are added, the resulting quadruplet augments the set to the 19-vertex rule. Therefore 15-vertex rule is a rotationally asymmetric reduction of the 19-vertex rule. In the natural 9-dimensional basis of the neighboring edge pairs one can define the R-matrix. It is indicated on the right: black entries for the 15-vertex rule and additional four gray ones for the 19-vertex one.

In this study we treat 15-vertex configuration simply as combinatorial objects and do not attempt to interpret them physically. For this particular reason we do not here pay attention to vertex weights. Furthermore we confine ourselves to a simple square domain with a given boundary condition (arrow/blank edge arrangement on the boundary). The reason is that we want to see if and how the limit shape phenomena, observed both in six-vertex and in 19-vertex models, arises in this \lq\lq interpolating model\rq\rq. The much more developed theory for dimer-type models (e.g. \cite{KO}) of course covers these domains and thereby perhaps gives some useful insight and perspective for these more involved models.

\vskip .2truein
\noindent In summary the bounded 15-vertex rule exhibits a far richer limit shape behavior even on a square domain than the six-vertex rule. Without any parametrization of the dynamic rule (corresponding at equilibrium to specific static weights), the mere fact that we can go beyond the Domain Wall Boundary Condition on the square gives a multitude of possibilities to shape the typical equilibrium configurations. Our simulations suggest a hierarchy in the complexity of the limits shapes appearing as a function of distance from DWBC.

\section{Preliminaries}

For the purposes of this study we concentrate on the square lattice ${\rm {\bf Z}}^2$ alone and a square domain in it. This is mainly to make our results comparable to a string of earlier studies in the same set up (e.g. \cite{CP}). Rules of our general type presupposing an even vertex degree of the ambient graph are possible on more exotic graphs/lattices and some of them have indeed been investigated earlier (see e.g. \cite{E3}). In spite of their undeniable interest, to keep the focus on the 6/15/19-vertex rule frame, we skip possible lattice dependent considerations here.

The {\bf Domain Wall Boundary Condition} (DWBC) is shown on the inscribed square on the left of Figure 2: the arrows are alternatively all in or all out for an entire side as we circumscribe the square. The diamond around the square will be motivated shortly.

\setcounter{figure}{1}
\begin{figure}[H]
\centerline{\includegraphics[height=4cm]{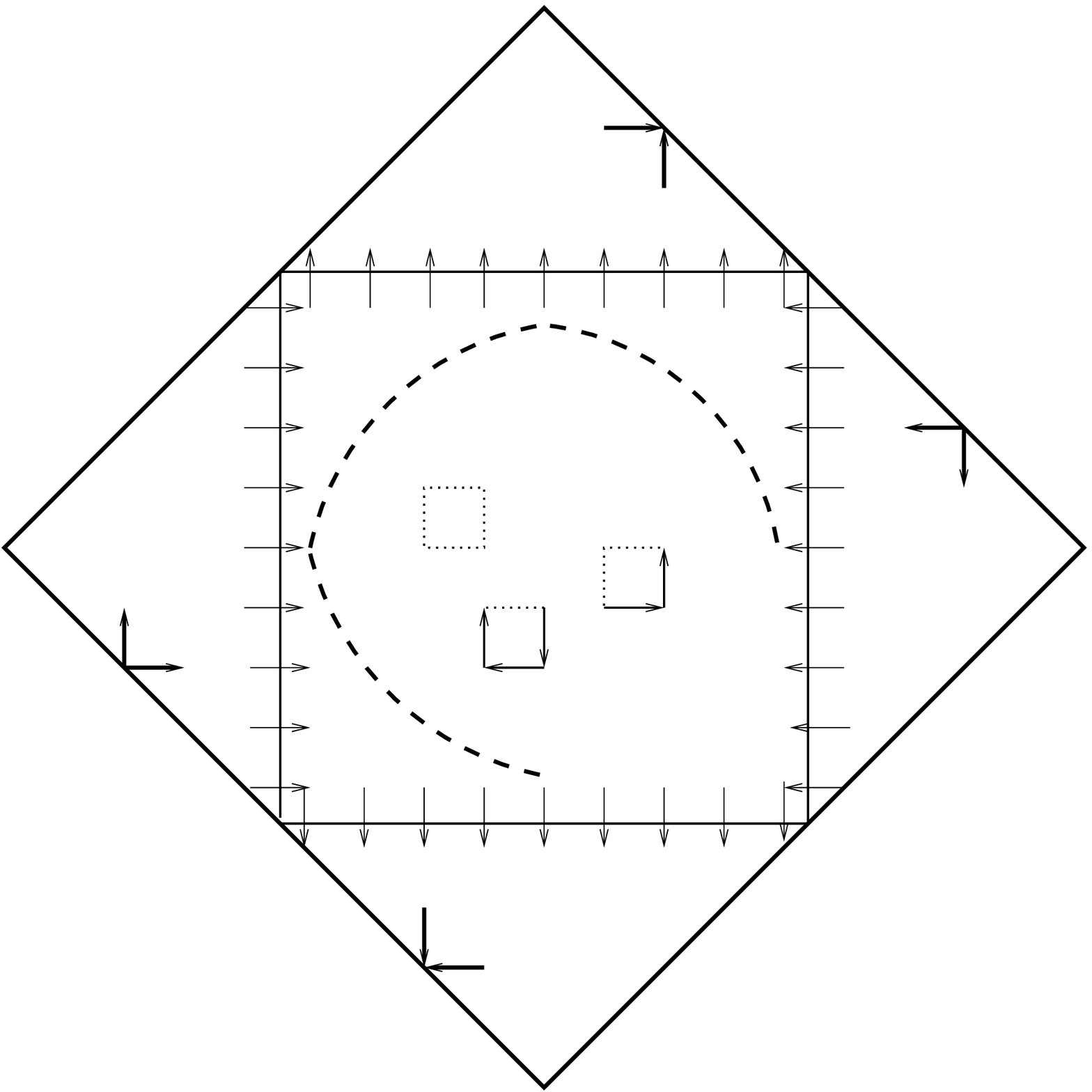} \hskip 2cm \includegraphics[height=4cm]{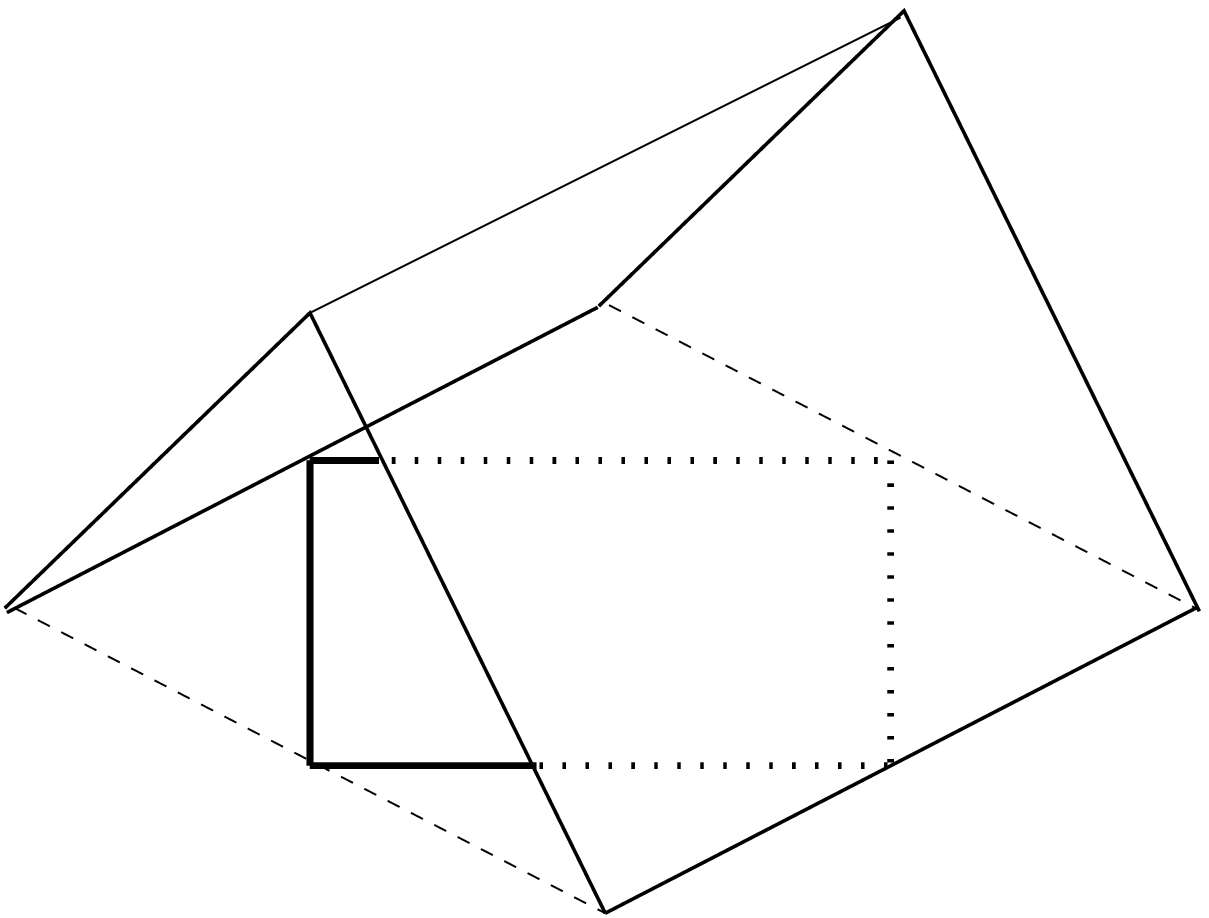}}
\caption{Domain wall boundary condition imposed on the inscribed square. A height surface.}
\end{figure}

\noindent \noindent {\bf Height} is a function from to the dual lattice ${\rm {\bf Z}}^2+\left(\frac{1}{2},\frac{1}{2}\right)$ to the integers. Moving from nearest neighboring dual lattice point to another the height increases by one if we cross a left-pointing configuration arrow, decreases by one if the crossed arrow is right-pointing and remains constant if no arrow is encountered. Given a configuration the height function defines a discrete surface over it which is unique up to an additive constant. The discrete derivative of height is referred to as the {\bf tilt} of the surface.

The indicated boundary arrow arrangement on the diamond of Figure 2 will force some of the configuration in the interior of the diamond and in particular projects down as DWBC on the boundary of the square. A particularly simple fill-in of this (smoothed) wire frame is on the right; height surface shaped like a {\bf ridge roof} with maximal tilts on both sides of the ridge.

\vskip .2truein
\noindent  Suppose now there is an unoriented interior edge $U$ in the domain. Whether it is horizontal/vertical correspondingly at its left/bottom end by the 15-vertex rule there must be a neighboring unoriented edge either towards W or S (or both). Hence there is an unoriented path to the boundary in the $3^{\rm rd}$ quadrant rooted at the left/bottom end of $U.$ Same argument connects the other end point of $U$ in its first quadrant with an unoriented path to the boundary, hence

\begin{prop}
  In a 15-vertex configuration unoriented interior edge implies unoriented boundary edges.
  \end{prop}


\noindent Since DWBC is fully occupied by arrows this obviously implies
\begin{cor}
  Under DWBC the sets of six- and 15-vertex configurations agree.
\end{cor}

\noindent DWBC is of course just one particular way of generating non-trivial limit shapes in the Ice-model context. Even sticking to a square domain, by setting the boundary arrow arrangement to have alternating extreme tilt ($\pm 1$) as one circumambulates the boundary yields a multitude of such examples (see \cite{E1}). But Corollary 1 implies that to find something genuinely novel in bounded 15-vertex model we need to go beyond DWBC and its fully arrow occupied relatives.

\section{Dynamic model}

%
%

\noindent To simulate the model we need a dynamic version of it. The {\bf elementary actions/moves} or {\bf flips} must be compatible with the set of static vertex rules. By their local nature we will have an efficient (even parallel) way of computing the perturbations of a given configuration in a bounded domain.

Since 15-vertex rule incorporates the six-vertex rule its well known generating action of reversal of unidirectional loops obviously has to be included (action II below). A bit more is needed to catch all of 15-vertex configurations. Figure 3 shows the set of minimal actions that connect all 15-vertex configurations sharing a boundary configuration.

\setcounter{figure}{2}
\begin{figure}[H]
\centerline{\includegraphics[height=1.75cm]{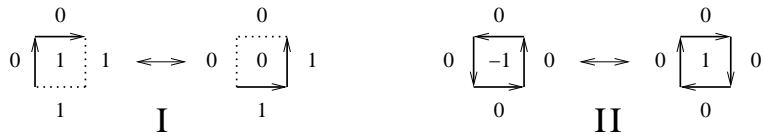}}
\caption{The local actions, I together with rotation by $\pi$. Nearest neighbor heights.}
\end{figure}

\noindent The unisotropy of the vertex rule (the framed center set in Figure 1) is reflected in the limited rotational symmetry of action I. One should also compare this action set to the much richer super set needed to generate the fully symmetric 19-vertex model (\cite{E2}).

To computationally best utilize these actions one splits the configuration on ${\rm {\bf Z}}^2$ into a checkerboard of even and odd unit squares each of which symbol value is determined by its edges. There are $81$ possible symbols according to the edge (un)orientations. For efficient computation the natural arrays of these units correspond geometrically to a diamond inscribing the square domain. In each iteration step all of the eligible even/odd squares are randomly updated. This can be done {\sl independently} of the other sites of the same color. After the iteration the odd/even lattice is updated according to the arrow (un)orientation changes. Our Probabilistic Cellular Automaton ((PCA) implements the action rules of Figure 3 in alternating sequence $\left({\rm I_e,\ II_e,\ I_o,\ II_o,\ I_e,\ldots}\right)$ with transition probabilities $1/2$ (for best speed) to each direction for both flip types I and II. It is simply a random walk on the finite graph of allowed configurations sharing the given boundary.

One of the diamond arrays has its boundary unit squares half fixed (two outermost edges) guaranteeing that the given boundary condition on the diamond is preserved. An example of this was indicated in Figure 2, left. The fixing of the diamond boundary in this given way forces the DWBC on the interior square for all times under the PCA.

\vskip .2truein
\noindent Since neither action changes the number of unoriented edges in the unit square we have for the 15-vertex rule:

\begin{prop}
  For a given boundary condition the number of unoriented arrows in the fill-in configurations is constant. 
\end{prop}

\noindent This is very different from 19-vertex model where the annihilation/creation of unidirectional unit loops is generically present. Then only lower and upper bounds for the arrow number are possible (\cite{E2}).

The following small observation will be useful in the next section. It is due to the fact that only action I ever moves a blank path of minimal width.

\begin{cor}
  An straight unoriented path from one side of the diamond across to the opposite one cannot branch under the 15-vertex dynamics.
  \end{cor}

\section{Game of boundaries}

\noindent To investigate the possible novel limit shape behavior in the 15-vertex rule we now step beyond DWBC. This is possible since we can introduce blank (unoriented) edges on the boundary, hence to the initial condition of our PCA. We will show how to do this and the results in roughly increasing complexity.

An already rather diverse set of more general boundary conditions is obtained by perturbing just the cross section of the \lq\lq ridge roof\rq\rq\ over the diamond in Figure 2, right. The entire height surface is still determined by the SW-NE translation of the cross section. The simplest version of this process of this is just shaving the sharp ridge off, Figure 4, middle. We call the resulting height surface K-type: the initial condition has two neighboring blank (unoriented) lines (zig-zags really) running parallel to the SW-NE diagonal. This alters DWBC on the inscribed square by introducing two doublets of unoriented edges to its SW and NE corners. This non-DWBC boundary condition on the square will prevail for all times under the PCA iteration.

\setcounter{figure}{3}
\begin{figure}[H]
\centerline{\includegraphics[height=2cm]{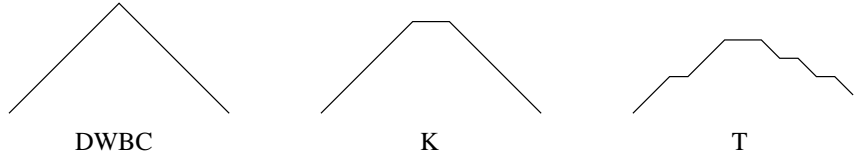}}
\caption{NW-SE sections of the height roofs of the initial conditions.}
\end{figure}

\noindent In the subsequent renderings of the simulations we concentrate of the cumulative action distributions of both types over the diamond. The reason for not showing snapshots of configurations is that action distributions are far more informative. From them one can at the equilibrium read the approximate limit shape, the dominant action creating it and the intensity variations within the active areas. In domino terminology, where there is no action, the configuration has a \lq\lq frozen region\rq\rq\ whereas the presence of either action indicates a \lq\lq temperate region\rq\rq. The intensity of the action at a given point (darkness of the pixel) reflects the local liveliness of typical configurations (number of close graph neighbors). These plots should be viewed as illustrations of the {\bf entropy geometry} of the given action under the fixed boundary condition on the diamond/square.

In the plots the diamond is tilted clockwise by $\pi/4$ from Figure 2 so that the imagined ridge of the height surface runs horizontally in a square. Neighboring plots have alternating slightly tinted backgrounds to better see the full diamond domain.

\vskip .2truein
\noindent Figure 5 shows the cumulative densities of actions I and II and their pointwise sum at the equilibrium for a K-type boundary/initial condition. With the given diamond size the boundary condition differs by 0.5\% from DWBC. The rendered distributions are shown within the entire diamond. The darker the distribution, the higher the activity. In the equilibrium outside the blob \& sickle there is no activity i.e. I+II defines the tempered domain.

As the PCA runs, the originally parallel neighboring blank zig-zags drift apart and action II takes over the center just like in the six-vertex model. By the last Corollary of the Section 3 the blank ribbons cannot branch and indeed they behave much like elastic bands attached by the initial condition to the center points of the sides. Action I on the ribbons feeds the other action by creating new unidirectional unit squares. Action II in turn is responsible for the bulk of disorder. In this interaction the blank frontiers are slowly pushed aside by the growing disk.

At the equilibrium the ribbons affect the limit shape only marginally. We expect that at the scaling limit their contribution is likely to be negligible and the six-vertex limit shape with uniform weights to prevail. Note also that this boundary condition is asymptotically DWBC. From Figure 5 it is nevertheless surprising that inspite of the strong intrinsic asymmetry in the 15-vertex Rule (and our initial state as well) even well before the scaling limit the shape of the tempered domain is remarkably symmetric.

\setcounter{figure}{4}
\begin{figure}[H]
\centerline{\includegraphics[height=4.5cm]{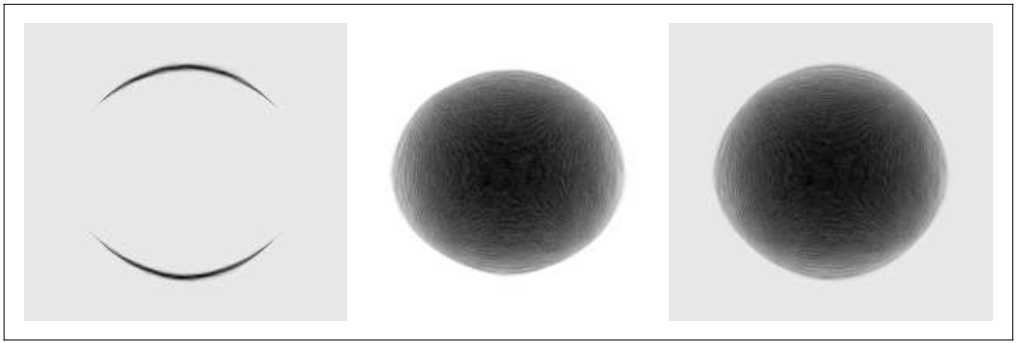}}
\caption{Equilibrium from K-type boundary condition. Densities of actions I, II and I+II (iterates 61-70.000, even sublattice within $206\times 206$ diamond, odd sublattice data is essentially indistinguishable).}
\end{figure}

 \noindent In addition to geometric matching the action density supports they agree well in intensity as indicated on the left of Figure 6 (same data, now sums rendered 3-d). Only with heavy bias can one distinguish their relative contributions near the boundary (right).

\setcounter{figure}{5}
\begin{figure}[H]
\centerline{\includegraphics[height=5cm]{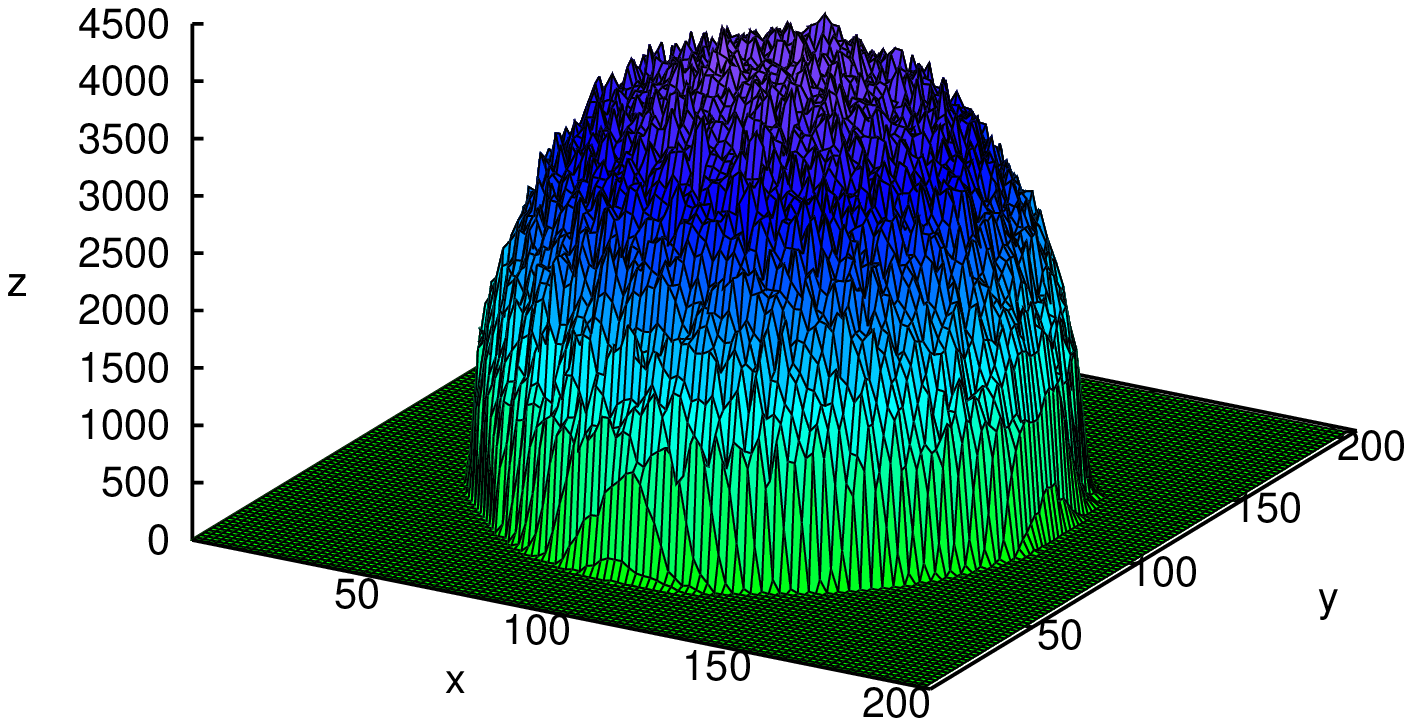} \includegraphics[height=5cm]{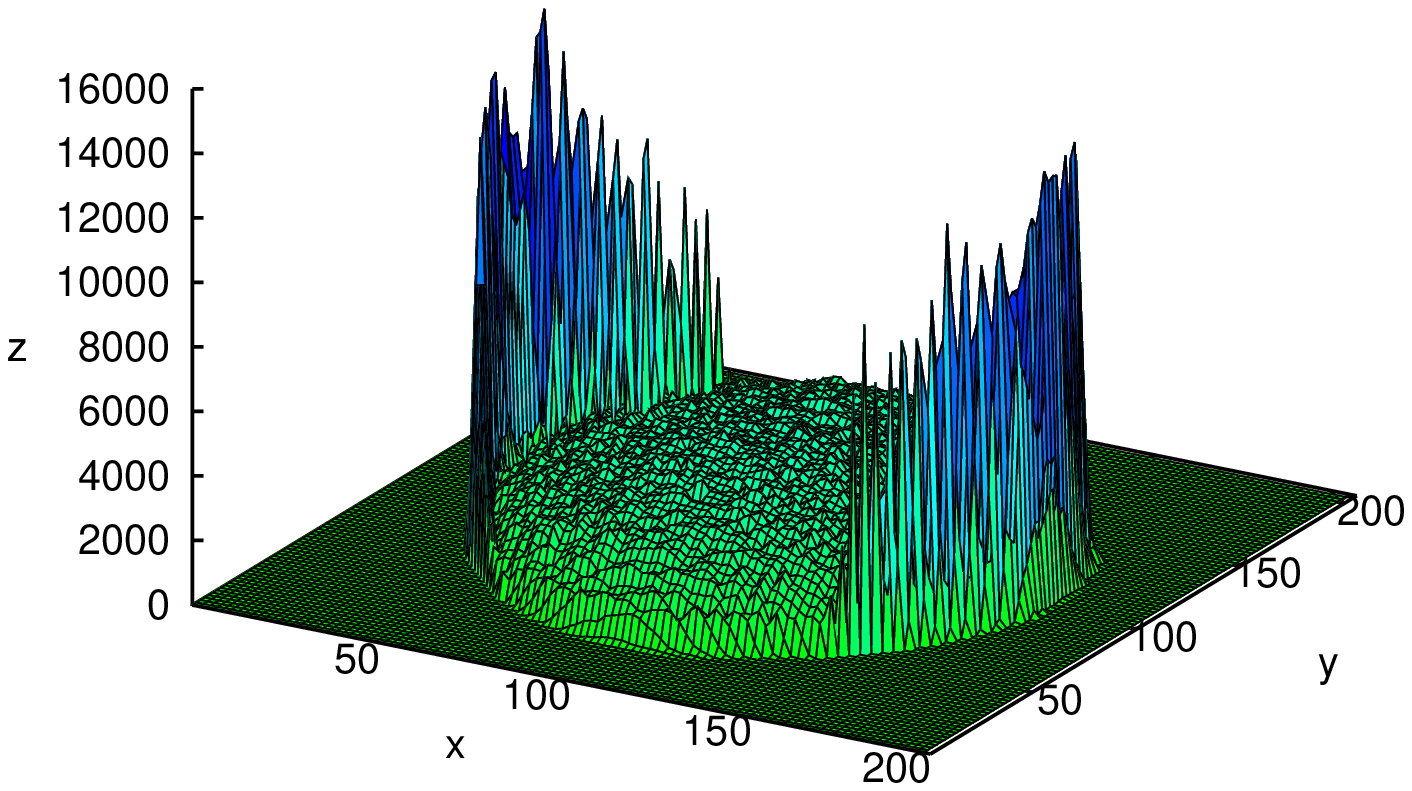}}
\caption{Actions superimposed: left, density sum, right: density sum with type I weighted tenfold.}
\end{figure}

\noindent If the blank paths (initially neighboring zig-zag-lines) move cleanly away from the top the limit shape forms in a best way as above. But the evolution of the shape even with K-type boundary can get somewhat exotic. If one of the paths entangles early on in the iteration at the top of the roof one may end up with a situation indicated in Figure 7 for an extended period.

\setcounter{figure}{6}
\begin{figure}[H]
\centerline{\includegraphics[height=3.5cm]{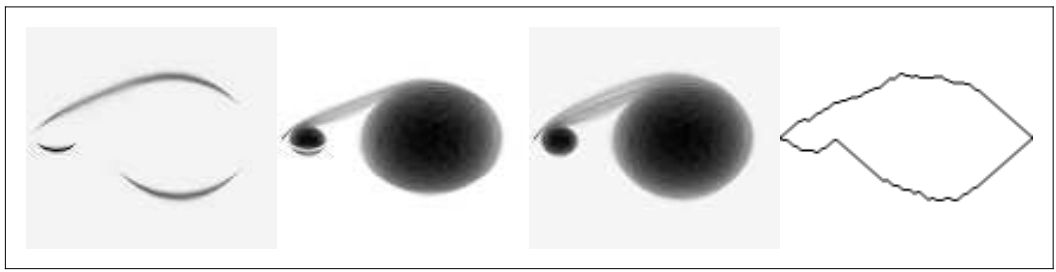}}
\caption{K-type boundary, unclean evolution. Cumulative actions I, II and I+II during iterates 40-60.000, $106\times 106$ diamond. Right: unoriented site locations only.}
\end{figure}

\noindent To further illuminate the state here, we have added a filtered snapshot at the end of the run (iterate 60.000) of the locations of vertices with unoriented incident edges (two per vertex). The corner point seems at least metastable. Whether it will always be removed by a large deviation event remains open. In that case the larger disk grows at the expense of the smaller resulting in a the typical limit shape of Figure 5.

\vskip .2truein
\noindent From the above one can perhaps guess that it might be possible to parallel transport copies of the blank SW-NE-diagonal path anywhere on the ridge roof and still get a 15-vertex legal initial condition. The cross section of such construction is on the right in Figure 4 and we call it T-type.\footnote{Just to continue with the ancient tradition going all the way back to dominoes and their Aztec-domain ([JPS]): K for Kheops/Khufu for its well known leveled top and T for Teotihuacan and its stepped/terraced pyramids.} These are of course even further away from DWBC than K-type. However they do not impose a constant boundary condition on the inscribed square anymore as we will shortly see.

Figure 8 illustrates typical equilibria in the case of T-type boundary condition. In the top row of the Figure in addition to the K-type flat top (two neighboring blank zig-zags) there are initially also two other blank zig-zags further out (\lq\lq terraces\rq\rq\ on the on the height surface). During the evolution they wiggle like a Brownian bridges but cannot match the action II driven central disordered expanding blob. They also can not cross each other oŕ the neutral paths pushed away from the diagonal. The limit shape seems to be a slightly oblate disk.

Note that the top and bottom zig-zags enter the inscribed square (inscribed diamond in the figures). Since they move around under the PCA iteration, there is no more a fixed boundary condition on the inscribed square. The natural domain for this type of boundary condition is the diamond itself, not the inscribed square.
 
\setcounter{figure}{7}
\begin{figure}[H]
  \centerline{\includegraphics[height=3.5cm]{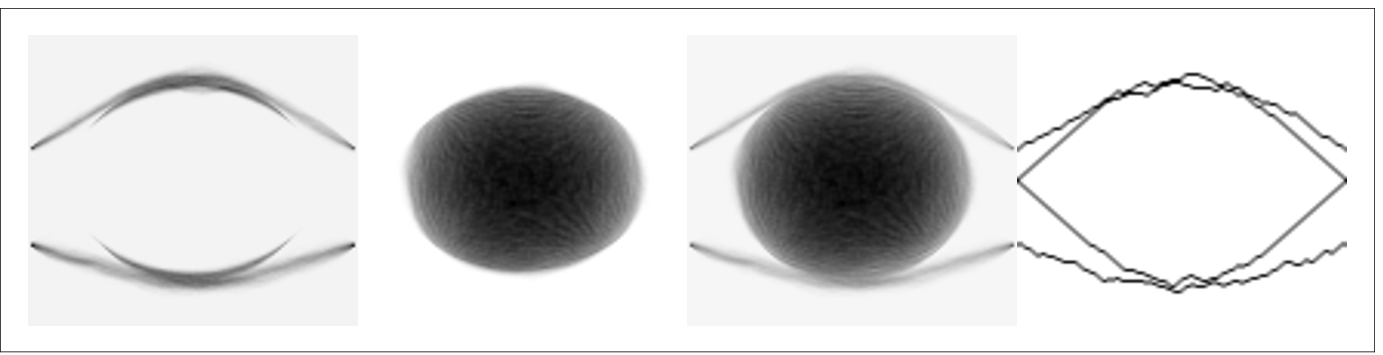}}
  \centerline{\includegraphics[height=3.5cm]{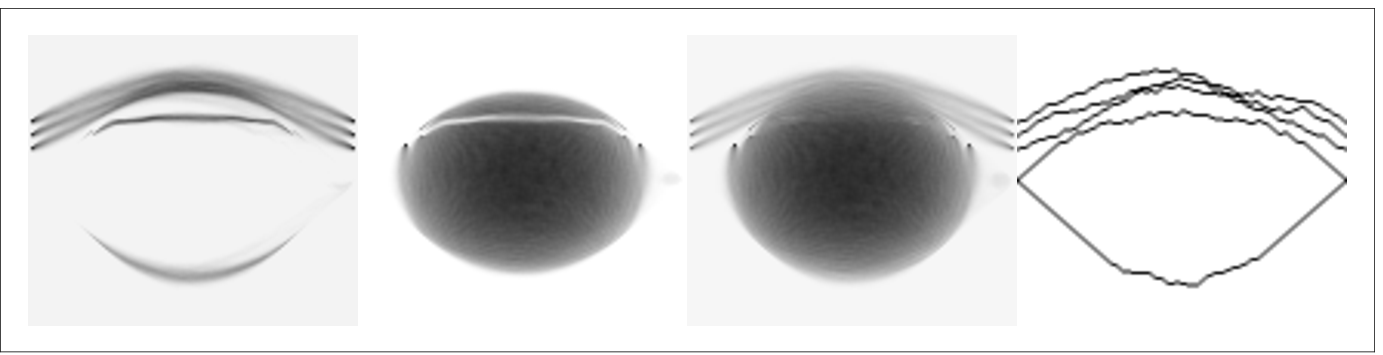}}
\caption{Equilibria from T-type initial conditions. Actions I, II and I+II. $106\times 106$ diamond. Top: 20-50.000 iterates, bottom 40-50.000 iterates. Right: unoriented edges at the end of each run.}
\end{figure}

\noindent Under still stronger boundary asymmetry one can encounter situations like the one rendered in the second set of Figure 8. Now in addition to the zig-zag pair at the roof top there are initially three more parallel blank zig-zags across one of the slopes. While slowly moving up in relative unison they heavily limit the motion of the blank ribbon sandwiched between them and the expanding central disk. The resulting equilibrium geometry seems to depend primarily on the relative distance of the attachment points of the blank ribbons on the sides.

\vskip .2truein
\noindent Finally Figure 9 shows a typical sample from a set of runs in a complicated situation where multiple zig-zags are initially bundled together at the top of the height surface (i.e. the original ridge of Figure 2 has been heavily planed). In this particular case there were initially six neighboring straight parallel zig-zags at the top. Their end points of course stick and the ribbons survive all times, but as they wiggle down the slopes they also interact intensely. And being confined to a narrow shuttle they cross the mid-ridge with high probability. This leads to a more complex case of the phenomenon already recorded in Figure 7. The resulting multi blob picture may well be generic. One possible scenario is a necklace of even size disks (equilibrium through surface tension which can be defined for the height surfaces) or in a (rare?) case a unique disk that has wiped out all the smaller ones. Possible metastability of all of this needs further study since already modest size runs indicate that at least the smallest disks can vanish in the PCA evolution.

\setcounter{figure}{8}
\begin{figure}[H]
  \centerline{\includegraphics[height=3.5cm]{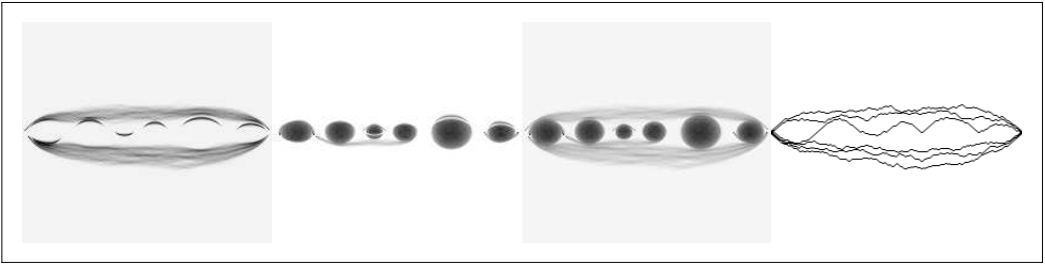}}
\caption{T-type initial condition with six neighboring blank zig-zags at the top. Actions I, II, I+II and unoriented at the end of the run. Even sublattice on a $206\times 206$ diamond, iterates 30-40.000.}
\end{figure}

\section{Conclusion}

\noindent 15-vertex model of Statistical Mechanics is in a way an interpolating model between the intensely studied six-vertex model and the more general but rather intimidating 19-vertex model. While 15-vertex model lacks the high complexity of the latter, the missing symmetry in the local rule sets it apart from both of the other models. We find that this strong microscopic unisotropy does not need to carry over to macroscopic features like limit shapes. Interestingly the bounded 15-vertex rule allows non-DWBC yet still under such boundary produces six-vertex like highly isotropic limit shape. Moreover there is a hierarchy of more complicated 15-vertex legal boundary conditions and non-trivial limit shapes seem possible and even likely in all of them. The characterization of them is challenging due to possible metastability effects.

\section{Acknowledgment}

\noindent The author would like to thank Nicolai Reshetikhin for bringing this model to his attention.

\end{document}